\documentclass[aps,pra,showpacs,amsmath,amssymb,amsfonts,lengthcheck,twocolumn,longbibliography,superscriptaddress]{revtex4-2}

\usepackage{epsfig,graphicx,graphics,amsmath,amssymb}
\usepackage[T1]{fontenc}
\usepackage[latin9,utf8]{inputenc}
\usepackage{xcolor}
\usepackage{amscd}
\usepackage{bm}
\usepackage{bbold}
\usepackage{psfrag}
\usepackage{mathrsfs}
\usepackage{booktabs}
\usepackage{svg}

\usepackage{graphicx}
\usepackage{braket}
\usepackage[colorlinks=true, allcolors=blue]{hyperref}

\usepackage{dsfont}
\usepackage{blindtext}

\newcommand{\bla}{bla\\bla\\bla\\bla\\bla}

\begin{document}

\title{Energy-error tradeoff in encoding quantum error correction}

\author{Josey Stevens}
\email{Josey.Stevens@jhuapl.edu}
\affiliation{Department of Physics, University of Maryland, Baltimore County, Baltimore, MD 21250, USA}
\affiliation{Quantum Science Institute, University of Maryland, Baltimore County, Baltimore, MD 21250, USA}

\author{Sebastian Deffner}
\email{deffner@umbc.edu}
\affiliation{Department of Physics, University of Maryland, Baltimore County, Baltimore, MD 21250, USA}
\affiliation{Quantum Science Institute, University of Maryland, Baltimore County, Baltimore, MD 21250, USA}
\affiliation{National Quantum Laboratory, College Park, MD 20740, USA}
\date{\today}

\begin{abstract}
While it has been widely recognized that genuine quantum advantage for practical problems might only be achieved with fault-tolerant quantum computers, it is still not entirely clear whether the required quantum error correction will be physically feasible. In the present work, we carefully analyze the required energy resources to encode the logical qubit states for repetition, perfect, and Steane codes. We find that there is a universal trade-off between the target precision and the required energetic resources. Importantly, we find that the energetic resources intimately depend on the specific physical realization of a quantum error correction code, and that the required resources scale exponentially with the targeted precision of the encoding.
\end{abstract}

\maketitle

\section{Introduction}

The development of practically useful quantum computers will require the realization of quantum error correcting (QEC) codes \cite{roffe2019quantum} in order to combat decoherence \cite{zurek2003decoherence,schlosshauer2019quantum} that is detrimental to the delicate nature of quantum correlations. Since it is exactly these quantum correlations from which quantum advantage can arise, realizing QEC might be the single most important challenge in the technological development of industry-scale quantum computers \cite{loepp2006protecting,Sanders2017}.
To date, hundreds of QEC schemes \cite{ErrorCorrectionZoo} have been proposed that differ greatly in their fault-tolerant implementation, scalability, size, and distance.
However, estimating the computational overhead required for QEC schemes is difficult due to challenges in simulating large quantum systems; yet, it is commonly expected that a single logical qubit will require hundreds to tens of thousands \cite{fowler2012surface,katabarwa2024early} of physical qubits.
This means that the vast majority of the computational complexity associated with quantum computing will be utilized in the overhead of QEC.

In realizing a quantum computer, designers will have to determine the resources required to operate and run their devices.
This will include overhead resources of materials and manufacturing costs as well as energy requirements for the sustainment of the coherent quantum system, i.e. maintaining vacuum, trapping ions, or cooling superconducting materials.
These costs are difficult to accurately project into the future, as they are subject to the unpredictability of technology advancement. Thus, the collected efforts to precisely quantify the energy requirements for quantum technologies has led to the foundation of a ``quantum energy initiative'' \cite{Auffeves2022PRXQ}.

However, energy is not only consumed in the technology surrounding the quantum processing unit, but also for the simple fact that any processing of information incurs a thermodynamic cost \cite{landauer1991PT}. Thus in a complete assessment of the resources required for quantum computing, will account for the cost of maintaining the computing environment as well as the resources required to perform the computation, namely the number of quantum gates and the energies associated with operating them.
Steps toward assessing the number of gates required for QEC exist in the form of threshold theorems \cite{knill1998resilient,aharonov1997fault} which provide a poly-log upper bound on the number of gates required to implement a QEC code provided a given error threshold is satisfied.

The primary purpose of this work is to analyze the energetic requirements for implementing QEC. Interestingly, it has already been recognized that the particular way any quantum algorithm is transpiled onto quantum hardware is of thermodynamic significance \cite{perlner2017thermodynamic,Buffoni2020QST,doucet2026thermodynamic}. However, a detailed looked a specific realizations of QEC appears to be lacking. Building off our recent work on the energetic requirements of single gate operations \cite{Stevens_2025}, in the present work we develop a comprehensive understanding of the energetic requirements for reptition, perfect, and Steane QECs.

To this end, in section \ref{Sec: Computational model} we introduce both the computational model, energy definitions, and noise types to be utilized throughout the treatment.
Section \ref{Sec: QEC} then provides an overview of QEC before exploring the tradeoff of energy and computation errors for the quantum repetition codes (subsection \ref{Subsection: QRC}) and the five qubit perfect code (subsection \ref{Subsection: 5QP}) and finally compare these with the Steane code (subsection \ref{Subsection: D3C}).
Particular attention is paid on exploring how changing the encoding circuits of a QEC affects the error performance.  

\section{Computational model}\label{Sec: Computational model}

Since our goal is to explore both the lower bound resource requirements and effects of gate-level noise on error corrected computations, we need a qubit control scheme that includes both a noise model and provides insight into the energetic requirements required to implement a gate. 
To this end, we utilize the precision-noise model previously developed only recently in Ref.~\cite{Stevens_2025}, in which we demonstrated that the lower-bound on the energy in the classical electromagnetic field used to control a quantum system is inversely related to the square of the gate-errors.
These errors arise from quantum fluctuations in the coherent electromagnetic field and allow us to analyze the energy-error tradeoff with true gate-level noise model that does not rely on insertion of random Pauli-errors.
Since classical magnetic and electric fields serve as the foundational control for trapped ion \cite{pogorelov2021compact,debnath2016demonstration}, semiconducting \cite{koppens2006driven,veldhorst2015two,pla2012single}, and superconducting qubits \cite{Li_2021,li2019manipulation,raftery2017direct,bardin2021microwaves}, these models provide insight into idealized implementations of these quantum computing platforms.   

In this error scheme, a quantum gate, $U_G$, is implemented efficiently using a constant Hamiltonian \cite{Aifer_2022} via a tensor-product decomposition of the form,
\begin{equation}
H_G = \frac{i}{\hbar\tau}\text{ln}(U_G) =     -i \sum_i \lambda_i \bigotimes_{j=1}^N C_{i,j}
\end{equation}
where $C_{i,j}$ are Hamiltonian terms that act on individual degrees of freedom and satisfy $\left[\bigotimes_{j=1}^N C_{i,j},\bigotimes_{j=1}^N C_{k,j}\right] = 0 $, and $\tau$ is the period over which the gate is driven.  
This control scheme allows for the formulation of a tight lower bound on the control-energy required to implement the gate,
\begin{equation} 
\label{Eq Ind Energy Bound combined}
    \braket{E_L^G}\ge \frac{\hbar \omega_0}{4}\sum_i \frac{\left|\lambda_i^G\right|^2}{\epsilon_i^2}.
\end{equation}
where $\omega_0$ is the lowest frequency in the control field, and $\epsilon_i$ is the standard deviation of the associated coefficient $\lambda_i$ resulting from the quantum fluctuations in the classical electromagnetic control field modes that couples to their respective term in the interaction Hamiltonian.
The bound is tight for all Hamiltonians of this form and it is a universal lower-bound for gates provided that a Hamiltonian is decomposed into an efficient minimal coefficient multiset, $\lambda^G$.  
A list of efficient coefficient multisets and lower bounds on the required control energy for common one and two-qubit gates is provided in Tab.~\ref{Gate List}.

\begin{table}
\centering

\begin{tabular}{ccc}
\toprule
Gate \qquad & $\lambda^\text{G}$ \qquad & $\frac{\braket{E_L^\text{G}}}{\hbar \omega_0}\,\left[\frac{\pi^2}{\epsilon^2}\right]$ \\
\midrule
X & $\{\pi/2,\pi/2\}$ & $1/8$ \\
Y & $\{\pi/2,\pi/2\}$ & $1/8$ \\
Z & $\{\pi/2,\pi/2\}$ & $1/8$ \\
C-X & $\{2\cdot\pi/4,2\cdot-\pi/4\}$ & $1/16$ \\
C-Y & $\{2\cdot\pi/4,2\cdot-\pi/4\}$ & $1/16$ \\
C-Z & $\{2\cdot\pi/4,2\cdot-\pi/4\}$ & $1/16$ \\
H  & $\{\pi/2,\pi/2\}$ & $1/8$ \\
Q & $\{\pi/2,2\cdot-\pi/\sqrt{8}\}$ & $1/8$ \\
S & $\{\pi/2\}$ & $1/16$ \\
\bottomrule
\end{tabular}
\caption{Table of common quantum gates, and the coefficient multisets that implement them efficiently, and the resulting lower bound on the control-energy needed to implement them.  X, Y, and Z are the Pauli spin operators, and C- are their associated controlled variants.  H is the Hadamard gate, $\text{Q} = (\text{Z}+\text{Y})/\sqrt{2}$, and S is the $P(\pi/2)$ phase shift gate.
Note the use of multiset notation where $\{N\cdot\lambda\}$ indicates $N$ independent but repeated inclusions of $\lambda$.}
\label{Gate List}
\end{table}

\begin{figure} 
    \centering
    \includegraphics[width=.98\linewidth]{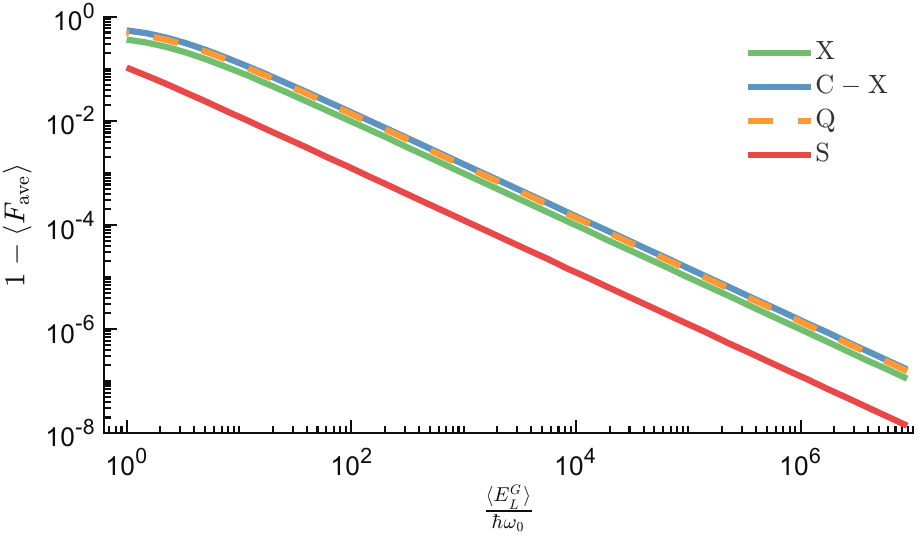}
    \caption{The gate error (one minus the average gate fidelity) as a function of the lower bound on the control energies required to implement the gates.  Shown here are the Pauli-$X$ gate (green), controlled-NOT gate (blue), $Q = \left(X + Y\right)\sqrt{2}$ (gold-dashed), and the phase gate, $S = \sqrt{Z}$ (red).
    Not shown here are the remaining Pauli gates and their controlled variants which show identical behavior to the Pauli-$X$ and controlled-NOT gates respectively.  For all gates analyzed the gate error is inversely proportional to the control energies for sufficiently high energies.}
    \label{Fig: Fidelity Energy}
\end{figure}

In the following analysis, we will utilize this noise model explicitly: gates are implemented via their most efficient Hamiltonian decomposition with noisy coefficient values drawn from  the normal distribution $\mathcal{N}( \lambda^G_i,\epsilon)$.
In this way, each gate can be understood as driving over or under rotations along the axis of their most efficient paths in the $n$-dimensional Bloch space.
We make the additional assumption that all gate coefficients regardless of their type (gate), qubit type (computational or ancilla), or location in the circuit are subject to the same noise distribution.
Further, we assume that all input states are in the computational basis state $\ket{0}$ and measurements are perfect projections on the computational basis without error or assessed energetic cost. 

While this model provides us immediate insights into the relationship between the control energy and the dynamical error in implementing the gate, in its current from this gives relatively little insight into the errors of our implemented gates.
To explore this, we will utilize the average gate fidelity \cite{PhysRevA.60.1888,NIELSEN2002249,PEDERSEN200747} between the desired gate, $U_G$ and the realized gate, $U_G'$ which is given by,
\begin{equation}
     F_{\text{ave}}(U,U') = \frac{1}{n\left(n + 1\right)}\left(n + \left|\text{Tr}\left(U^\dagger U'\right)\right|^2\right),
\end{equation}
where $n$ is the dimension of the Hilbert space.
By performing simulations of noisy gates and taking the ensemble average, $\left<F_\text{ave}\right>$, in Fig.~\ref{Fig: Fidelity Energy} we see that at high energies, the average gate error scales as the inverse of the control energy, i.e., $ 1-\left<F_\text{ave}\right> \propto \left<E_L^G\right>^{-1}$.

Since we aim to study the energetic and error tradeoffs on not just state maintenance \cite{Acharya2025} but also real computation, we perform a single $X$-gate acting on the logical qubit. Real algorithms are comprised of many gates, however for the sake of clarity and tractability we restrict ourselves here to the minimal case of only a single gate.

\begin{figure}
    \centering
    \includegraphics[width=0.98\linewidth]{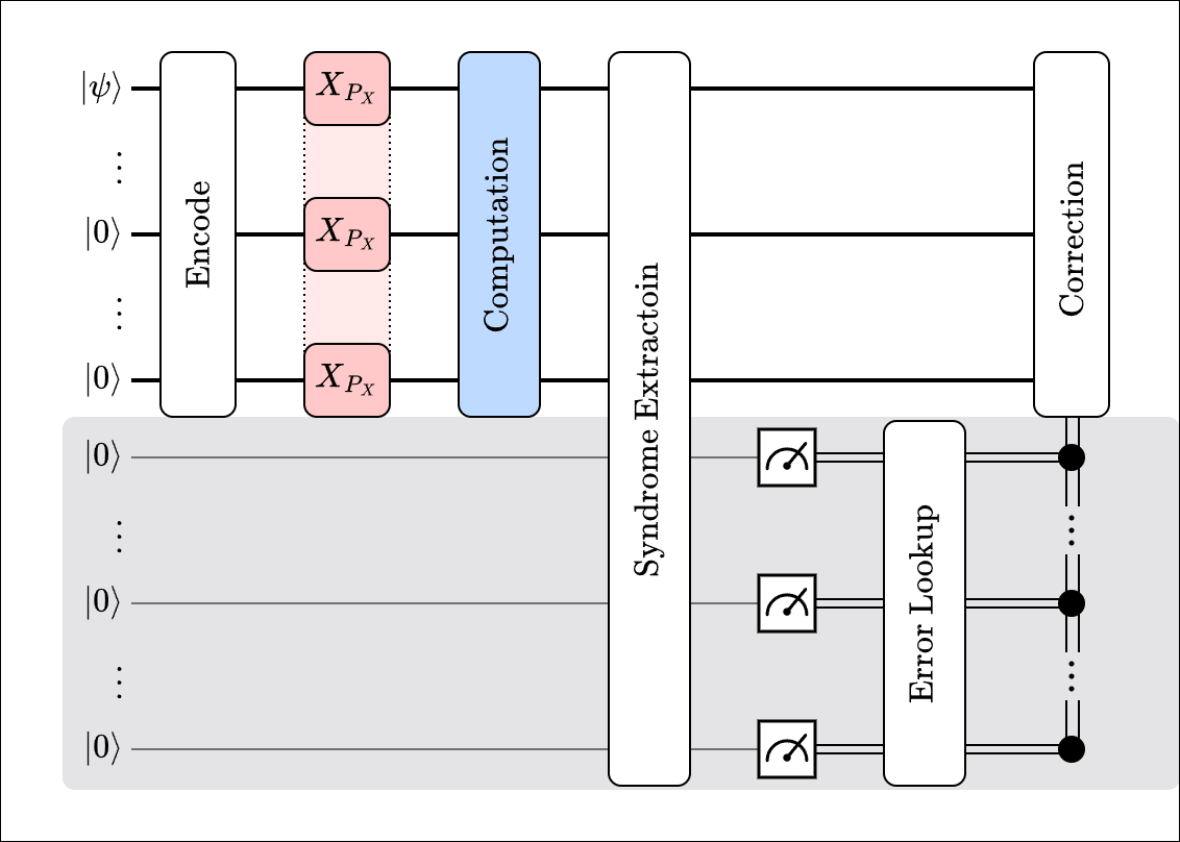}
    \caption{Generic depiction of the QEC process utilized in this analysis described here.  The channel errors are highlighted in red, computation on the logical qubit is highlighted in blue, and all explicit QEC steps are shown without color in greyscale.  The ancilla qubit registers are highlighted with gray and classical registers are shown as doubled bars.}
    \label{Figure: Generic Computation}
\end{figure}

In addition to the gate level noise, we subject our computation to a channel level noise model where each computational qubit is subject to an independent probabilistic bit-flip with probability $P_X$.
This channel error occurs immediately after the encoding block is completed.
We do not include the effects of channel noise on the ancilla qubits.
A generic representation of this computation is provided in Fig.~\ref{Figure: Generic Computation}.
In all investigations, we will characterize the effective performance of our computation by calculating the average error of the projective-Z measurements, $\langle Z \rangle_{error}$, relative to their expected value in an error-free computation.

\section{Quantum Error Correcting Codes} \label{Sec: QEC}

To analyze the bounds on the energetic requirements for quantum error correction, we first must choose the appropriate codes, determine circuits for their implementation, and finally bound their control energy by counting the gates according to Tab.~\ref{Gate List}.

One aspect of QEC which has been under-explored is the fact that functionally equivalent (but distinct) quantum circuits lose their equivalency in the presence of either gate or channel level noise since the noisy processes generally do not commute with the multiple circuit representations.
In the following discussion, we address this aspect directly by analyzing the error correcting performance of multiple encoding circuits for both repetition codes and the five qubit perfect code.

\subsection{Quantum repetition codes} \label{Subsection: QRC}

The simplest set of QEC methods are the $N-$qubit repetition codes \cite{nielsen2010quantum,Wootton2018PRA,Dulong2026PRB}.
Unlike many QEC codes that correct for arbitrary error, quantum repetition codes only protect the logical qubit against errors along a specified axis.

In general, an $N$-qubit repetition code can protect against $N/2-1$ errors in the logical subspace.
This is accomplished by preparing the logical qubit as $\ket{\psi_\text{L}} = \alpha_0\ket{0...0} + \alpha_1\ket{1...1}$ by entangling the input qubit with $N-1$ qubits via controlled-NOT gates.
Logical gates are transversal, i.e. $\text{X}_\text{L} = \text{X}...\text{X}$, thus our computation requires $N$ X-gates.
Syndrome extraction is carried out through parity measurements; in the stabilizer framework, their generators are compactly expressed as $\text{g}_i=\text{I}_1...\text{Z}_i \text{Z}_{i+1}...\text{I}_N$, requiring $2(N-1)$ controlled-NOT gates.

In our error model, in the absence of gate errors ($\epsilon\rightarrow0$), the $N$-qubit repetition code has a failure rate given by,
\begin{equation}\label{Eq: Failure Rate}
\begin{split}
   F(P_X,N)&= 1-\sum _{k=0}^\frac{N-1}{2} \frac{N! P_X^k (1-P_X)^{N-k}}{k! (N-k)!} \\ &\approx\binom{N}{(N+1)/2}P_\text{X}^{\frac{N+1}{2}} ,
\end{split}
\end{equation}
in leading order of $P_X$.
This result is obtained by describing the number of simultaneous errors with a binomial distribution and noting that any number of bit flips less than $\left(N-1\right)/2$ will be corrected. 

After extraction of the syndrome, any detected errors must then be corrected, requiring, on average
\begin{equation}
\sum _{k=0}^{\frac{N-1}{2}} \frac{k N! P_X^k (1-P_X)^{N-k}}{k! (N-k)!} \approx P_X N
\end{equation}
NOT-gates, in leading order of $P_X$ which follows analogously to the derivation of Eq.~\ref{Eq: Failure Rate}.

Since the true average number of corrections is dependent on gate noise ($\epsilon>0$) and the number of correction gates can never exceed $(N-1)/2$, we will neglect the energy required to perform the error correction.
In Tab.~\ref{Table: Repetition Codes List}, we collect the gate counts and calculate the total minimum control-energy for the $N$ qubit repetition codes.
We observe that in the worst case, the energy required for correction of the detected errors cannot exceed $1/5$ of the total control-energy budget already accounted for. 

\begin{table}
\centering
\begin{tabular}{cccc}
\toprule
ECC & X Count & C-X Count & $\frac{\braket{E_L^\text{ECC}}}{\hbar  \omega_0}\,\left[\frac{\epsilon^2}{\pi^2}\right]$ \\
\midrule
Bare Qubit & 1 & 0 & $1/8$ \\
3-Qubit Repetition & 3  & 6 & $3/4$\\
5-Qubit Repetition & 5 & 12 & $11/8$\\
7-Qubit Repetition & 7 & 18 &$2$\\
$N$-Qubit Repetition & N & 3(N-1) &$(5N-3)/16$\\
\bottomrule
\end{tabular}
\caption{Summary of gate types, counts, total control-energy require to implement the $N$-qubit repetition codes.}
 \label{Table: Repetition Codes List}
\end{table}

As we wish to analyze the resource to performance tradeoffs of our error correcting scheme, we have minimized the impact of several factors that plague real QEC efforts.  
Namely, we have assumed that both initial qubit state preparation and projective measurements are performed perfectly with no (assessed) energetic cost, our channel errors are unitary, uncorrelated, and occur at only one point in our quantum circuit (immediately after an encode).

\subsubsection{Logical equivalency of encoding circuits}

In almost all cases, a given unitary can be implemented with any number of logically equivalent circuits.
Choosing between these different representations is often dependent on properties of the computing device (native gates or connectivity topology) or either spatial or time complexity considerations \cite{holmes2020impact,cruz2019efficient}.
In addition to these explicit or resource constraints are considerations based on errors.
These have been explored in the context of both dynamic \cite{murali2019noise} and static \cite{Maldonado2022} gate decompositions.
Here we will explore an additional noise based consideration based on the simple observation: the noisy realizations of equivalent circuits are not necessarily equivalent themselves and explore how this impacts the performance of the repetition codes.

\begin{figure}
    \centering
    \includegraphics[width=0.98\linewidth]{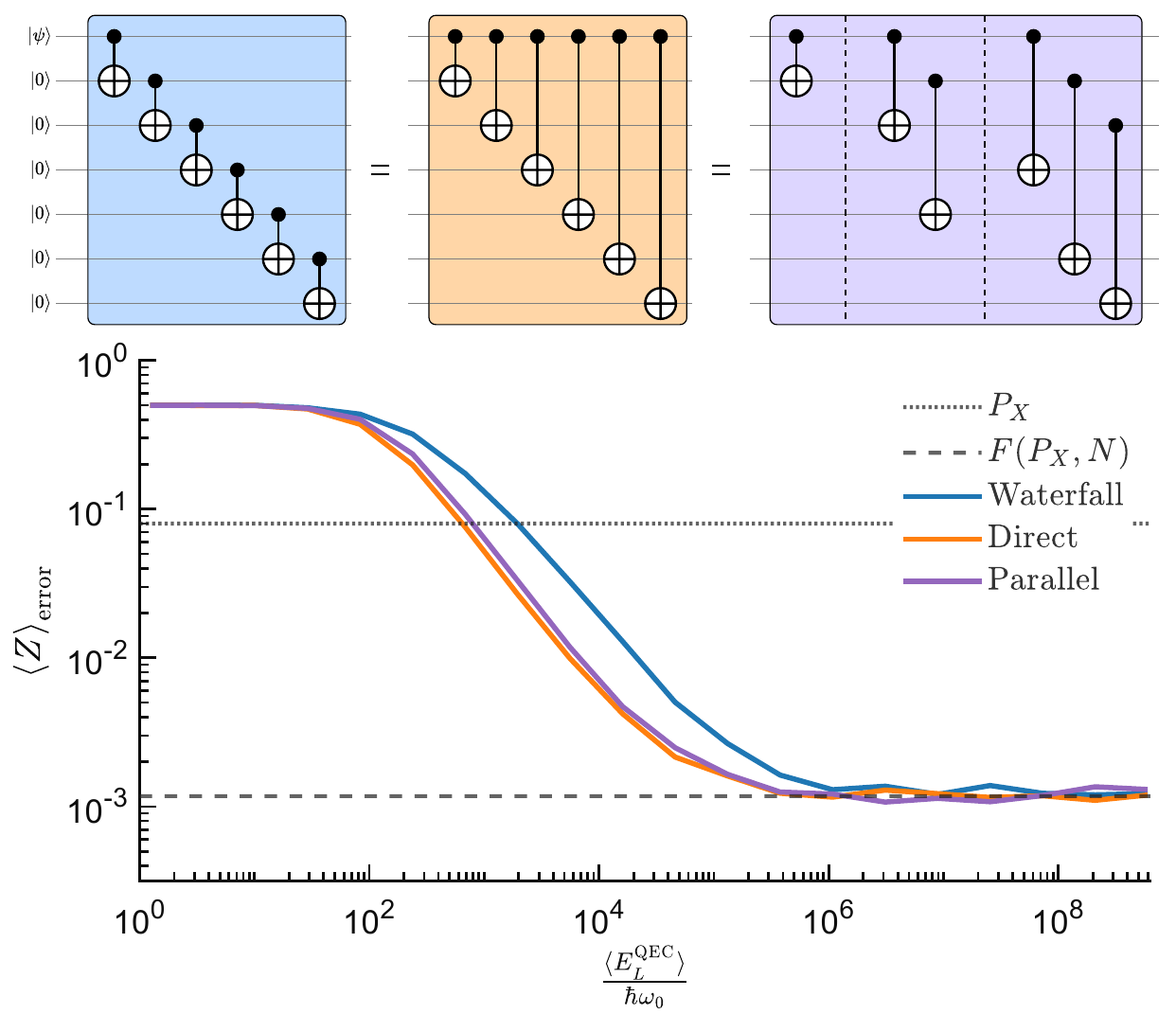}
    \caption{Computation error as a function of algorithm control-energy for the $N$-qubit repetition code with three different encoding schemes with a fixed channel noise of $P_X=0.08$.  
    The waterfall (blue), direct (orange), and parallel (purple) encoding circuits demonstrate similar performance at both the low and high energy regimes but demonstrate markedly different error rates at intermediate energy levels, with the direct encoding demonstrating the best performance.
    For the parallel encoding we have grouped portions of the circuit that can be executed in parallel.}
    \label{fig: 7 Qubit Encoding}
\end{figure}

In Fig.~\ref{fig: 7 Qubit Encoding} we show three circuits for state encoding for the $7-$qubit repetition code. 
We first note that these circuits are only logically equivalent on the subspace of inputs appropriate for encoding, but they are not equivalent if arbitrary input states are allowed.
In the first method, the input information cascades from one qubit to its neighbors in a method that resembles a waterfall.
The second method takes a different approach where the input qubit is repeatedly utilized to directly impart its information on all other qubits.
The third method breaks up the encode into time slices such that in each slice the maximum number of C-NOT gates are utilized via parallel implementation.
For the remainder of this work we will refer to these encoding methods as \textit{waterfall}, \textit{direct}, and \textit{parallel} encoding respectively.

Waterfall and direct encoding are connectivity driven, i.e., the waterfall encoding is the only option available for a nearest neighbor connected device while direct encoding can be utilized on an all-to-all connected device.
In contrast, parallel encoding minimizes the complexity of the encoding time and allows implementation in logarithmic time of the code size \cite{cruz2019efficient}. 

We observe that, despite all three encoding circuits requiring the exact same number and types of gates, the QEC protected computation exhibits significantly different performance as a function of the control energy between the three different encodings.
Namely, direct encoding performs the best with parallel encoding showing slightly higher error and the waterfall encoding demonstrating significantly higher error rates.

This demonstrates how device connectivity can be a major limiting factor for QEC performance even when transpilation does not add a large overhead from a gate-counting perspective. For the remainder of this section we will utilize direct encoding to compare between the $N-$qubit repetition codes.

\subsubsection{Repetition code scaling}

\begin{figure*} 
    \centering
    \includegraphics[width=1\linewidth]{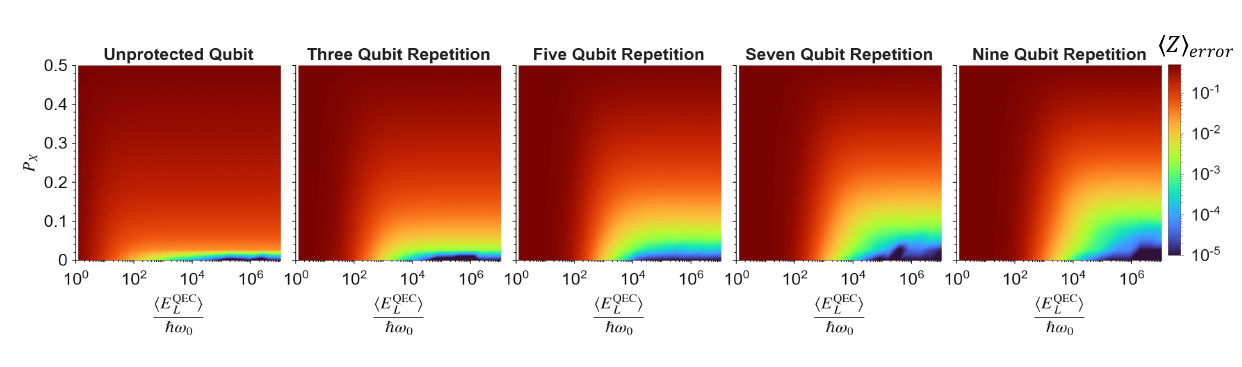}
    \caption{Error rates for the $N$-qubit repetition code as both a function of the circuit control-energy and the channel error rate.
    Generally, the larger distance codes provide more robust protection at sufficiently high energy regimes, but enter those regimes at increasingly high energy.}
    \label{Fig: Repetition Code Surface Plots}
\end{figure*}

Having established how the gate counts and energetics scale as the size of the repetition code and having established how the gate fidelities scale as a function of energy, we can turn to our primary exercise: assessing how the energetic requirements of the $N-$qubit repetition codes scale as a function of their size.
This is explored in Fig.~\ref{Fig: Repetition Code Surface Plots} where we show the error rates for the $N-$qubit repetition codes a function of both $P_X$ and control-energy.

In general, we observe that, as expected, if the gate fidelity is high, the larger codes provide significantly better protection against high channel noise.
However, we also observe that the larger codes are more susceptible to gate errors and require increasingly high control-energy to enable this performance.
These results are intuitive since the larger codes are subject to more noisy gates at all stages of the QEC protected computation.

\begin{figure}
    \centering
    \includegraphics[width=1\linewidth]{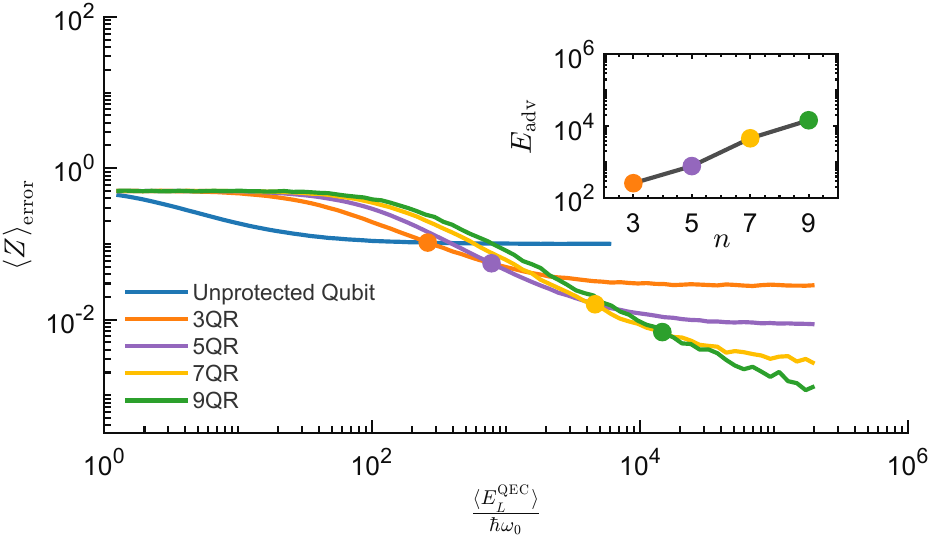}
    \caption{Error rates of computation a computation subject to a fixed channel rate of $P_X = 0.10$.  
    The computation is carried out on the bare qubit (blue) and circuits protected by the three (orange), five (purple), seven (yellow), and nine (green) qubit repetition code.  Energies where the $N$-qubit repetition code reach an advantage over the $(N-1)$-qubit repetition code are highlighted with colored circles.
    The inset plots these energies versus the size (distance) of the code and demonstrate exponential scaling behavior.  
    }
    \label{fig:EnergyScalingRepetition}
\end{figure}

We explore the relationship between code size and energetic requirements more closely in Fig.~\ref{fig:EnergyScalingRepetition} where we show how (for a fixed channel error rate) the logical error scales as a function of code size.
We observe that the control energy required for the $N-$qubit code to overtake the $(N-1)-$qubit code scales exponentially with the scaling parameters dependent on the channel error.

\subsection{Five Qubit Perfect Code} \label{Subsection: 5QP}

The five qubit perfect code $[[5,3,1]]$ \cite{PhysRevA.54.3824,laflamme1996perfect} is the smallest QEC code that can correct an arbitrary single qubit error, including phase rotations which are outside of the computational and error basis assessed in our computational model.
The cost of this robustness to expanded channel error is not only an increase in the number of channel qubits, but also more complicated logical state preparation, decoding, and syndrome extraction circuits. Where the repetition codes allowed for simple local readout and classical counting to measure the logical qubit, to do a measurement on the perfect code we run the encoding circuit \textit{backwards} to turn the logical qubit into a single physical qubit and perform measurements on the physical qubit.
    
\begin{figure}
    \centering
    \includegraphics[width=1\linewidth]{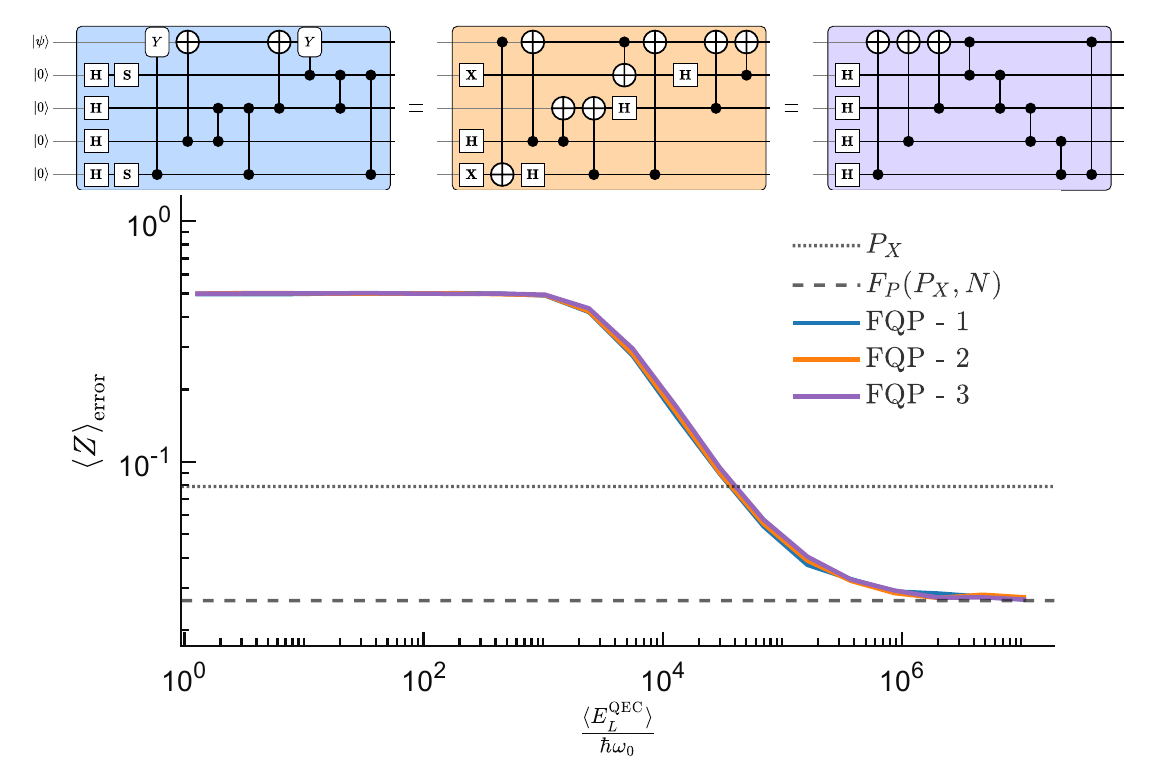}
    \caption{Average error rate of the computation protected by the five-qubit perfect code with various encoding schemes subject to a constant channel error rate of $P_X = 0.08$.  The encoding circuits (above) are from \cite{hsieh2006analytical} (blue), \cite{mondal2024quantum} (orange), and \cite{de2016universal} (purple).
    The three encoding circuits demonstrate similar error rates as a function of circuit control-energy.  
    }
    \label{fig:FQR}
\end{figure}

    In Fig.~\ref{fig: 7 Qubit Encoding} we explored how the different encoding circuits for the repetition codes can drive significantly different error behaviors. Similarly in Fig.~\ref{fig:FQR} we explore three different encoding circuits for the perfect code.
    We observe that there are minor differences in the error behaviors between the three encoding circuits, but do not see the major trends observed for the repetition codes.
    From these results, it is unclear if a significantly improved encoding scheme for the perfect code exists, or how implementations limited by device topology may affect these results.  
    Due to the similarities in all three encoding schemes, for remaining sections we will utilize only the first encoding in Fig.~\ref{fig:FQR}.

\subsection{Distance-3 codes} \label{Subsection: D3C}

Finally, we explore the energetic-error tradeoff between the three-qubit repetition,  perfect, and Steane codes~\cite{PhysRevLett.77.793} each of which encodes one logical qubit in three, five, and seven qubits, respectively, and are all capable of correcting a single error from our noisy channel.

For the Steane code $[[7,1,3]]$, we employ the encoding circuit from Ref.~\cite{mondal2024quantum} which utilizes only Hadamard and controlled-NOT gates for the encoding and decoding process.
Like the perfect code, the Steane code is able to correct any arbitrary single qubit error on the channel. However, unlike the perfect where every syndrome utilizes all of the computational qubits, the Stean code admits stabilizers that utilize at most four code qubits.

\begin{figure}
    \centering
    \includegraphics[width=1\linewidth]{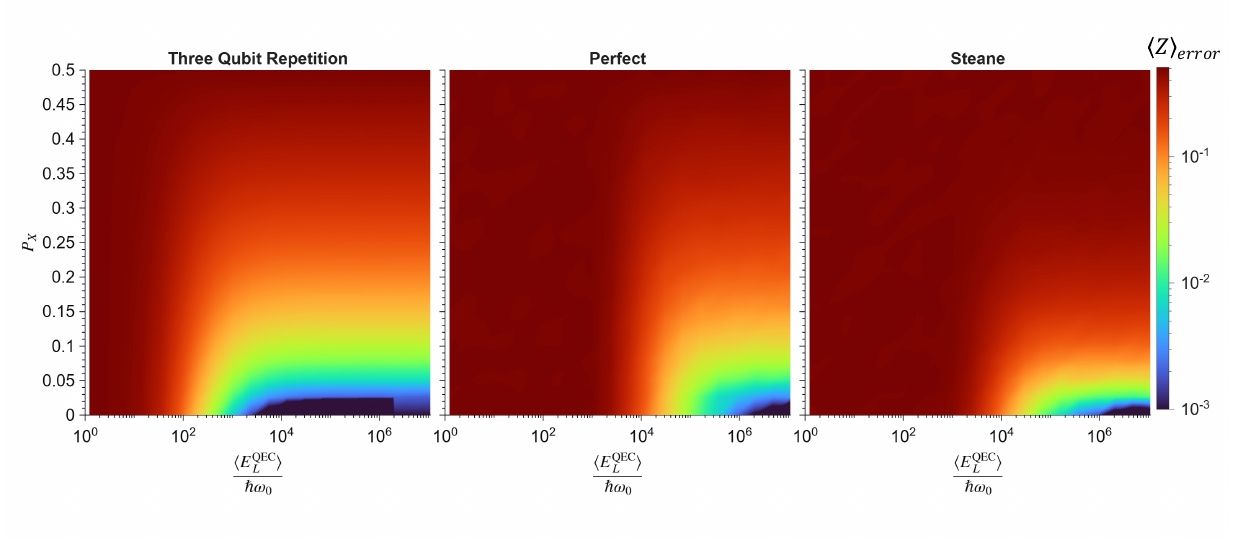}
    \caption{Computation error rates for distance three codes demonstrated by the three qubit repetition, five qubit perfect, and (seven qubit) Steane code (ordered left to right) as both a function of the circuit control-energy and the channel error rate.
    We generally see that the constant-distance code with the fewest number of code qubits provides the most robust protection for a given channel error rate and the complexity of the encoding circuit is a driving factor of the energy required to achieve protection.}
    \label{fig:CompareDistance3CodesHeatmaps}
\end{figure}

\begin{figure}
    \centering
    \includegraphics[width=1\linewidth]{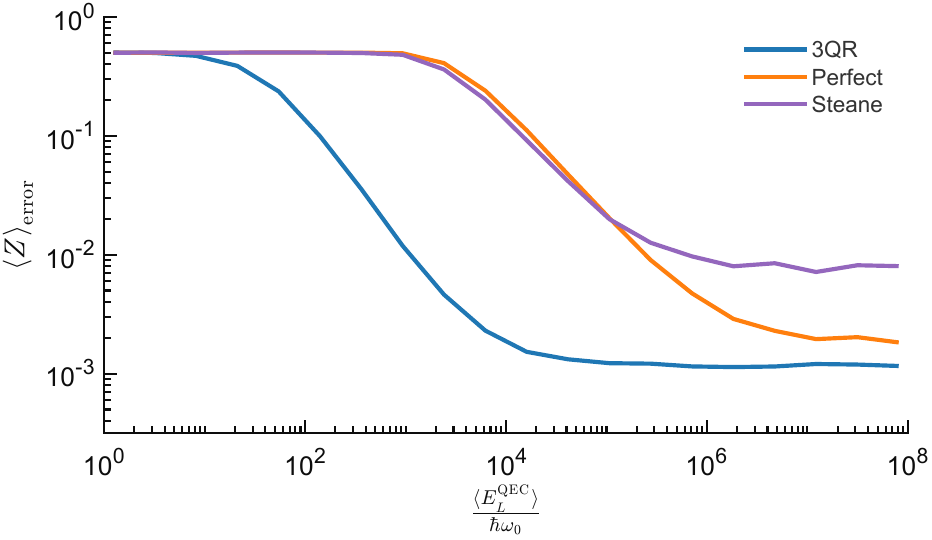}
    \caption{Computation error rate as a function of the circuit control energy given a fixed channel error rate of $P_X = 0.02$ for computations protected by the distance-three QEC.  
    The three-qubit perfect code (blue) achieves the best overall performance at all energy levels while the five qubit perfect code (orange) demonstrating the second best error rate at high energies and the (seven) qubit Steane code demonstrating the worst protection at high energies but achieving an onset of error protection at lower energies than the perfect code.}
    \label{fig:CompareDistance3Codes}
\end{figure}

In Fig.~\ref{fig:CompareDistance3CodesHeatmaps} we see that the three qubit repetition code requires significantly less energy before it begins achieving error correction relative to the more complex codes and at sufficiently high energies, provides significantly more robust protection.
By examining Fig.~\ref{fig:CompareDistance3Codes} we conclude that despite the Steane code's larger size and complex encoding scheme, it achieves error correcting behavior at slightly lower energies than the perfect code but is quickly overtaken due to the fact that the perfect code is subject to less overall channel noise.

In these results, we see that unlike the repetition codes which admitted straightforward scaling assessments, assessing the control-energy requirements for more increasingly complex codes of fixed distance is non-trivial.
The control energy requirements for very large codes will likely be driven by code distance, sparsity, and device topology.
Additionally, the number of encoding circuits for large codes will grow rapidly due to combinatorial growth in different encoding schemes and logically degenerate circuits that implement those schemes.

\section{Concluding remarks}

A key requirement in developing practical quantum computers is assessing the resource requirements needed to realize them.
Toward that goal, in this work, we have performed numeric simulations of small scale QEC utilizing a gate-level noise model that relates gate-errors to the control-energy.

We also demonstrate how noisy implementations of logically equivalent encoding circuits can have a large impact on the performance of a QEC code.
We observe that generically more complex QEC codes require more control energy to achieve error correction and observe that repetition codes of increasing distance require exponentially larger amounts of energy to reach error correcting advantage.

These observations lead to many natural questions which may be crucial to the realization of fault-tolerant quantum computers.
Investigations may be needed to determine if QEC codes of increasing distance generically require exponentially increasing energy, and if so, what design choices via complexity tradeoffs \cite{Stevens_2025} might be incorporated to reduce this cost.

In order to best utilize noisy quantum hardware, optimal encoding circuits for a specific QEC with noisy gates will need to be explored alongside the development of gate topologies.
Finally, the methods utilized in this analysis can, in principle, be applied to many aspects of quantum computing from a hardware and algorithmic viewpoint in order to identify the resource costs that must be addressed in the development of practical quantum computers.

\acknowledgments{We thank Martin B. Plenio and Gregory Quiroz for insightful conversations. S.D. acknowledges support from the John Templeton Foundation under Grant No. 63626.}


\appendix

\section{Fault Tolerant Measurements}

The discussion in Sec.~\ref{Sec: QEC} did not attempt to add fault-tolerance to stabilizer measurements.
To supplement this gap, this Appendix augments both the error and energy perspectives.
Namely, we find that not only is the added energetic overhead of fault tolerant measurement process not offset by an associated decrease in error, but under this idealistic noise model considering only gate precision errors, the addition on fault-tolerant processes increases overall error at fixed energy.

We proceed by utilizing the methods outlined in Ref.~\cite{gottesman2024surviving} utilizing fault tolerant cat state preparation and associated readout.
For the $N$-qubit repetition codes, $N$ two-qubit cat states are needed, each requiring a single Hadamard gate and one controlled-NOT for the initial preparation and an additional two controlled-not gates for each round of state validation.

Once the cat states are prepared an additional $2N$ controlled-NOT gates and $N$ Hadamard gates are needed to perform the fault-tolerant measurement. 
This brings the total added lower bound energetic requirement for fault-tolerant processing to at least $\braket{E_L^\text{FT}}/({\hbar \pi^2 \omega_0}) = N(7+2v)/16$ where $v$ is the number of validation rounds performed.

Even though this figure does not include the energy needed to prepare a cat state if an error is detected in validation, it adds a considerable amount to the overall energy budget for the overall QEC protocol, when a single validation round is used, the overall energy budget is increased by a factor of 3.25 when $N=3$ and 2.8 when $N\rightarrow\infty$.  

With the energetic overhead of fault-tolerant measurement assessed, we can now turn toward the error assessment.
We use the same error models as in Sect.~\ref{Sec: Computational model} where channel errors do not apply to the ancilla qubits utilized for syndrome extraction.

\begin{figure} 
    \centering
    \includegraphics[width=0.98\linewidth]{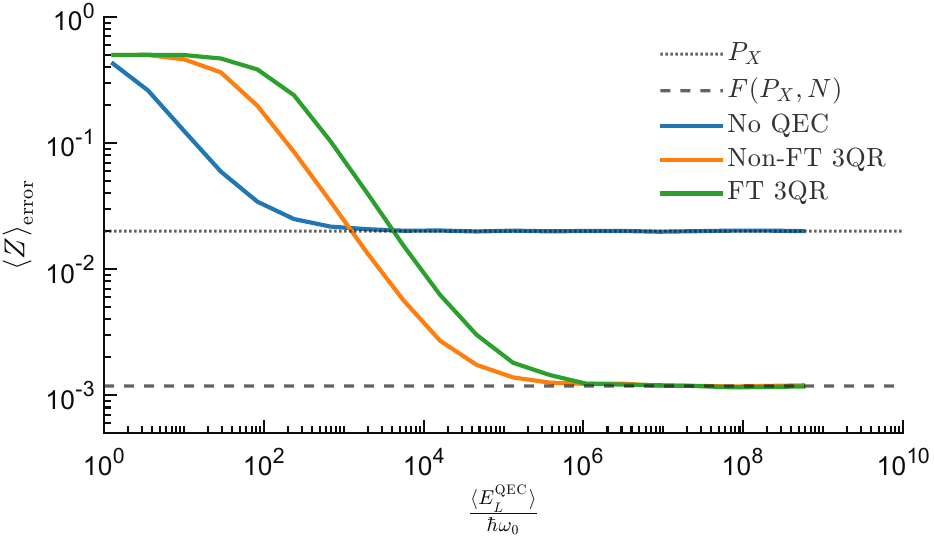}
    \caption{Computation error rate of a the bare qubit (blue) three qubit repetition code without fault-tolerant readout (orange) and with fault-tolerant readout (green) for a fixed channel error rate of $P_X = .02$. 
    The fault-tolerant code utilized here uses a single round of validition for the GHZ-state preparation i.e. $v=1$.
    Both repetition codes saturate their expected protection rates at high energies but the fault-tolerant circuit requires significantly higher control-energies to achieve robust protection.}
    \label{Fig: Fault Tolerant}
\end{figure}

In this case, as is shown in Fig.~\ref{Fig: Fault Tolerant}, we observe that the fault-tolerant measurement processes increase the overall error of the algorithm at a given control-energy.
Both the non-FT and FT implementations approach the theoretical performance at large energies. 
Indeed, if ancilla channel errors and other errors in the measurement process were included, we would see more improvement from the FT implementation. However, since our goal is to establish lower bounds on the control-energy and error relationship, and all FT measurement processes aim to restore noisy measurement processes to their non-noisy ideal realizations with large energetic overheads, the assessments in Sec.~\ref{Sec: QEC} serve as lower bounds as intended.  

\bibliography{0_references}

\end{document}